\documentclass[aps,prl,twocolumn,amsmath,amssymb]{revtex4}
\usepackage{times}
\usepackage{graphicx}
\usepackage{amsfonts}
\usepackage{amsmath, amsthm, amssymb}
\usepackage{dsfont}
\usepackage{color}
\newcommand{\e}[1]{\mathrm{e}^{#1}}
\newcommand{\eg}{\textit{e.g. }}%[syn: f.eks., for example, for instance]
\newcommand{\etal}{\emph{et al.}}
\def\i{\mathrm{i}}                            

\newcommand{\g}{\underline{\gamma}}
\newcommand{\gt}{\underline{\tilde{\gamma}}}
\newcommand{\N}{\underline{\mathcal{N}}}
\newcommand{\Nt}{\underline{\tilde{\mathcal{N}}}}

\newcommand{\Au}{A_{\uparrow}}
\newcommand{\Ad}{A_{\downarrow}}
\newcommand{\tAu}{\tilde{A}_{\uparrow}}
\newcommand{\tAd}{\tilde{A}_{\downarrow}}
\newcommand{\tAt}{\tilde{A}_t}
\newcommand{\tAs}{\tilde{A}_s}
\newcommand{\Bu}{B_{\uparrow}}
\newcommand{\Bd}{B_{\downarrow}}
\newcommand{\tBu}{\tilde{B}_{\uparrow}}
\newcommand{\tBd}{\tilde{B}_{\downarrow}}
\newcommand{\tBt}{\tilde{B}_t}
\newcommand{\tBs}{\tilde{B}_s}

\begin{document}

\title{Spin supercurrent and phase-tunable triplet Cooper pairs via magnetic insulators}

\author{Ingvild Gomperud and Jacob Linder}

\affiliation{Department of Physics, Norwegian University of
Science and Technology, N-7491 Trondheim, Norway}

\begin{abstract}
We demonstrate theoretically that a dissipationless spin-current can flow a long distance through a diffusive normal metal by using superconductors interfaced with magnetic insulators. The magnitude of this spin-supercurrent is controlled via the magnetization orientation of the magnetic insulators. The spin-supercurrent obtained in this way is conserved in the normal metal just like the charge-current and interestingly has a term which is independent of the superconducting phase difference. The quantum state of the system can be switched between 0 and $\pi$ by reversing the insulators from a parallel to antiparallel configuration with an external field. We show that the spin-current is carried through the normal metal by superconducting triplet (odd-frequency) correlations and that the superconducting phase difference can be used to enhance these, leaving clear spectroscopic fingerprints in the density of states.
\end{abstract}
 
\date{\today}

\maketitle

\section{Introduction}
Using superconductors as active components in spintronics devices is a research field that has attracted increasing activity in recent years \cite{linder_nphys_15}. Such a synergy becomes possible both due to the special behavior of spin-polarized quasiparticles in superconductors, featuring extremely long spin lifetimes and spin relaxation lengths \cite{yang_nmat_10, quay, hubler}, and because superconducting Cooper pairs can become spin-polarized \cite{bergeret_prl_01, kadigrobov_epl_01, eschrig_prl_03}. This type of Cooper pairs occur not only in superconductors with intrinsic triplet pairing, such as UGe$_2$ \cite{saxena_nature_00} and its cousins URhGe and UCoGe \cite{aoki}, but can in fact be artificially engineered in hybrid structures between conventional superconductors and magnetic materials \cite{bergeret_rmp_05, buzdin_rmp_05}. For samples with substantial impurity scattering, which is the experimentally most common scenario, these triplet Cooper pairs acquire a special property known as odd-frequency symmetry \cite{berizinskii} in order to satisfy the Pauli principle. What this means is while the Cooper pair wavefunction is symmetric under both an exchange of space- and spin-coordinates, it is antisymmetric under an exchange of time-coordinates. This property leads to remarkable features such as gapless superconductivity \cite{balatsky}, anomalous Meissner effects \cite{tanaka_prb_05, yokoyama_prl_11, alidoust_prb_14, fominov_prb_15}, and the possibility to create spin-supercurrents in diffusive ferromagnetic materials \cite{eschrig_phystoday_11}. It is worth mentioning that paramagnetic Meissner effects and zero-bias conductance peaks can also originate from other types of effects which are not related to unconventional superconductivity, as shown previously in the context of $d$-wave superconductivity in the cuprates \cite{kostic_prb_96, shan_prb_05, zhao_prb_10} and more recently for topological insulators \cite{levy_prl_13}. The quality of the materials and avoiding crashing the STM tip into the sample are of paramount importance for proper identification of unusual types of superconductivity such as the odd-frequency one mentioned above.

In the context of utilizing superconductors for spintronics purposes, the possibility of spin-supercurrents in ferromagnetic materials \cite{keizer} has earned the triplet Cooper pairs much attention. It is known that in structures featuring inhomogeneous magnetic order, such as intrinsically textured ferromagnets like Ho \cite{robinson_science_10, sprungmann}, or multilayers with several ferromagnets \cite{khaire_prl_10}, triplet supercurrents can arise even when using conventional $s$-wave superconductors which feature spinless Cooper pairs. However, it can be difficult practically to control the magnetization direction of each of the individual layers when using complex structures as in Ref. \cite{khaire_prl_10} to create the spin-supercurrent. A large amount of works have studied how triplet supercurrents can arise in various types of structures including both weakly and strongly polarized ferromagnets (see \eg Refs. \cite{zaikin_prb_09, houzet_prb_07, trifunovic_prl_10, fogelstrom_prb_00, shomali_njp_11, grein_prl_09, alidoust_prb_10, bobkova, liu, kupfer, iovan, klose, ostaay}). However, it would be highly desirable to create a spin-supercurrent flowing in a \textit{normal (non-magnetic)} metal with a minimum amount of magnetic elements required due to the ensuing simplification in how to control the existence, and the properties, of the spin-supercurrent.

\begin{figure}[b!]
\centering
\resizebox{0.5\textwidth}{!}{
\includegraphics{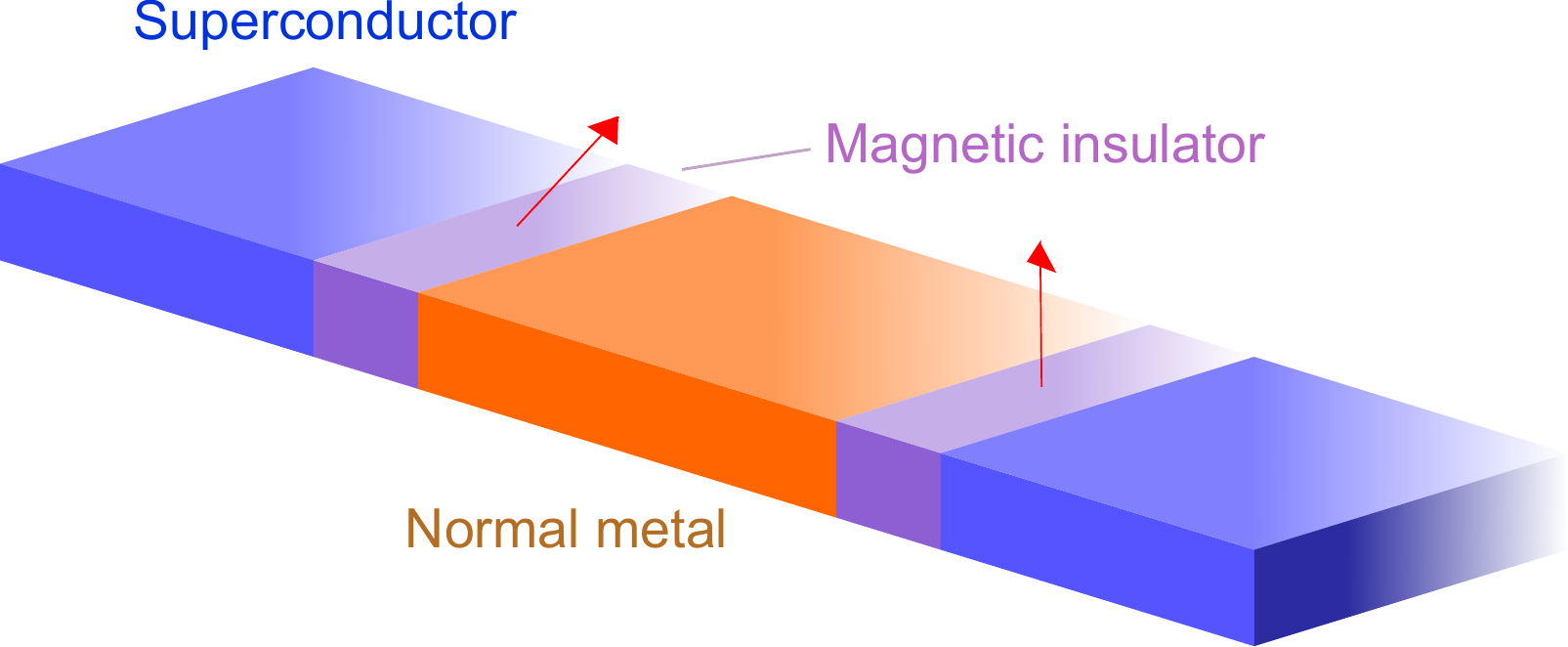}}
\caption{(Color online) The proposed setup: a Josephson junction with magnetic insulators (MIs) inserted between the superconductors and the normal metal. The magnetic insulators have a magnetization that due to shape anisotropy is expected to be confined to the plane perpendicular to the tunneling direction. The magnetic moments of the MIs on the left and right side of the junction, $\mathbf{m}_L$ and $\mathbf{m}_R$, may be misaligned and an applied superconducting phase difference across the junction drives both a charge and spin supercurrent.}
\label{fig:model} 
\end{figure}

Now, it was recently experimentally shown in Ref. \cite{li_prl_13} that by using magnetic insulators (MI) in a superconducting spin-valve setup, it was possible to control the superconducting critical temperature $T_c$ by changing the relative magnetization configuration from parallel (P) to antiparallel (AP) by application of an external field, causing an essentially infinite magnetoresistance effect. This finding prompted us to pose the question: what happens when magnetic insulators are used in a Josephson junction that has a normal, non-magnetic metal as its weak link? Can this create a spin-polarized dissipationless flow and, if so, can such a spin-supercurrent be controlled externally? We here demonstrate that the presence of magnetic insulators in a Josephson junction offers an interesting way to create a conserved spin-supercurrent in a normal metal (see Fig. \ref{fig:model}). This spin-flow is controlled with the relative misalignment angle between the magnetic insulators. A main advantage with this setup, compared to previous proposals using ferromagnets, is that the magnetic configuration is easily tunable since the magnetization of one single very thin (1-2 nm) magnetic insulator needs to be altered.

Moreover, the spin-supercurrent is carried by odd-frequency Cooper pairs which leave a clear spectroscopic fingerprint in the density of states. We show that by tuning the superconducting phase difference $\theta$, one can qualitatively change the nature of the proximity effect from a conventional singlet one to a triplet one \textit{in situ}. We also show that the presence of magnetic insulators not only creates a spin-supercurrent, but that it has important consequences for the quantum phase of the system which undergoes a dynamic 0-$\pi$ transition for a single sample by changing from a P to AP configuration for the insulators.

\section{Theory}

We use the quasiclassical theory of superconductivity \cite{rammer, chandrasekhar} in the diffusive limit, where the physics is described by the Green matrix function $\check{g}$ of the system which is an 8$\times$8 matrix in Keldysh-Nambu space. It is defined in terms of the retarded, advanced, and Keldysh part of the Green function: $\check{g} = \begin{pmatrix} 
g^R & g^K \\
0 & g^A \\
\end{pmatrix}.$ In the absence of non-equilibrium effects, such as applied voltages and temperature-gradients, it is sufficient to consider the retarded part $g^\text{R}\equiv g$, which may be parametrized conveniently as follows \cite{ricatti}:
\begin{align}
g &= \begin{pmatrix}
\N(\underline{1}+\g\gt) & 2\N\g \\
-2\Nt\gt & -\Nt(\underline{1}+\gt\g) \\
\end{pmatrix}, g^2 = 1.
\end{align}
We have defined $\N =(\underline{1}-\g\gt)^{-1}$ for normalization and the $\tilde{\ldots}$ operation means complex conjugation and reversal of quasiparticle energy. The Ricatti-matrices $\{\g,\gt\}$ are 2$\times$2 matrices in spin space and the Green function $g$ satisfies the Usadel equation \cite{usadel} in the normal metal 
\begin{align}\label{eq:usadel}
D\partial_x(g\partial_x g) + \i[\varepsilon\rho_3,g]=0.
\end{align}
Here, $D$ is the diffusion coefficient of the normal metal, $\rho_3=\text{diag}(\underline{1},-\underline{1})$, and $\varepsilon$ is the quasiparticle energy measured relative the Fermi level. In order to account for the magnetic insulators at the interfaces, we use spin-dependent boundary conditions discussed in Ref. \cite{cottet}. The most important effect of the magnetic insulators is the spin-dependent phase-shifts experienced by quasiparticles scattering at the interface, which are described by a parameter $G_\varphi$ to be defined below. The superconducting regions are described by bulk Green functions which for the left and right side of the junction are denoted $g_L$ and $g_R$, where 
\begin{align}
\g_j = \i\underline{\sigma}_y s/(\underline{1}+\underline{1}c)\e{\i\theta_j},\; \gt_j = -\i\underline{\sigma}_y s/(\underline{1}+\underline{1}c)\e{-\i\theta_j},
\end{align}
with $j=\{L,R\}$. We have introduced
\begin{align} 
s=\sinh\Theta,\; c=\cosh\Theta \text{, with } \Theta = \text{atanh}(\Delta_0/\varepsilon),
\end{align}
where $\Delta_0$ is the magnitude of the superconducting order parameter. The bulk solution is an excellent approximation for low interface transparencies when using large superconducting reservoirs. We have used the second Pauli matrix $\underline{\sigma}_y$, and the superconducting phase difference across the junction is defined as $\theta\equiv \theta_R-\theta_L$. The boundary conditions read:
\begin{align}
2 d \zeta_L g \partial_x g = [g_L,g] + \i G_\varphi^L[\mathcal{M}_L,g] \text{ at } x=0,\notag\\
2 d \zeta_Rg \partial_x g = [g,g_R] - \i G_\varphi^R[\mathcal{M}_R,g] \text{ at } x=d,\notag\\
\end{align}
where $\zeta_j = R_{B,j}/R_N$ is the ratio between the normal-state barrier resistance on side $j$ and the resistance of the normal metal, and $G_\varphi^j = -\sum_n\text{d}\varphi_n/\sum_n T_n$ where $T_n$ is the transmission probability for channel $n$ and d$\varphi_n$ are the spin-dependent part of the phase-shifts picked up by particles scattered at the interface. Finally, the matrix $\mathcal{M}_j$ describes the orientation of the magnetic moment of the magnetic insulator on side $j$, while $d$ is the length of the normal metal. Experimentally, it is likely that the magnetic insulators will have exchange fields lying in the plane perpendicular to the tunneling direction due to shape anisotropy if one uses a layered 'pancake' geometry for the junction. This case, and other configurations, are covered by us setting the right interface to  $\mathcal{M}_R = \text{diag}(\underline{\sigma}_z,\underline{\sigma}_z)$ whereas the left interface is allowed to have an arbitrary orientation, i.e. $\mathcal{M}_L = \cos\alpha\text{diag}(\underline{\sigma}_z,\underline{\sigma}_z) + \sin\phi\sin\alpha\text{diag}(\underline{\sigma}_y,-\underline{\sigma}_y) + \cos\phi\sin\alpha\text{diag}(\underline{\sigma}_x,\underline{\sigma}_x)$. Here, $\phi$ is the azimuthal angle in the $xy$-plane and $\alpha$ is the angle between the magnetization and the $z$-axis.  For later use, we define the magnetic moments of the insulators on the left and right side as $\mathbf{m}_L$ and $\mathbf{m}_R$.

The boundary conditions used here can also be extended \cite{cottet} to include a magnetoresistance term $G_\text{MR}$ which accounts for the different transmission probabilities for spin-$\uparrow$ and spin-$\downarrow$ particles. Inclusion of such a term amounts mainly to an overall reduction of the superconducting proximity effect and we have explicitly verified numerically that the spin-supercurrent exists even in its presence. The magnetic moment associated with each magnetic insulator should be understood as the net average moment of the interface region, since a disordered interface might have an internal magnetic structure. Moreover, interfaces in hybrid structures are intrinsically accompanied by the lack of inversion symmetry. For this reason, spin-orbit effects could arise at the interface and modify the spin-dependent scattering at the superconducting interface \cite{linder_prl_11, liu_prl_14, sun_prb_15}. However, it is currently unknown how to incorporate such interfacial spin-orbit scattering in the boundary conditions of quasiclassical theory. Nevertheless, even if such a mechanism existed, the spin-dependent phase-shifts due to the magnetic insulators captured by the parameter $G_\varphi$ in our work is sufficient to produce triplet Cooper pairs, and so we do not expect that a second mechanism that accomplishes the same thing (due to spin-orbit interaction) would bring about any major changes.

Finally, we will later on include non-ideal effects such as spin-flip scattering due to magnetic impurities and spin-orbit impurity scattering to see how they influence the spin- and charge-flow as well as the density of states in the system. These are accounted for \cite{rammer, chandrasekhar} by adding extra self-energy terms in the commutator part of the Usadel equation Eq. (\ref{eq:usadel}):
\begin{align}\label{eq:flip}
\begin{split}
&\text{Magnetic impurities:   } \Sigma_\mathrm{sf} = \frac{i}{8\tau_\mathrm{sf}}\boldsymbol{\tau}g\boldsymbol{\tau}, \\
&\text{Spin-orbit scattering:   } \Sigma_\mathrm{so} = \frac{i}{8\tau_\mathrm{so}}\boldsymbol{\tau}\rho_3 g\rho_3\boldsymbol{\tau},
\end{split}
\end{align} 
Here we have defined the matrix vector $\boldsymbol{\tau} = (\tau_x,\tau_y,\tau_z)$, where the components are given by:
\begin{align}\label{eq:tauNu}
\tau_\nu = \text{diag}(\underline{\sigma}_\nu,\underline{\sigma}_\nu^*),\; \nu=\{x,y,z\}.
\end{align}
For future reference, we introduce the normalized strength of magnetic impurity and spin-orbit scattering as $g_\text{sf} = 1/8\Delta_0\tau_\text{sf}$ and $g_\text{so} = 1/8\Delta_0\tau_\text{so}$, where $\tau_\text{sf/so}$ are the relaxation times associated with each type of scattering.

\section{Results}

\subsection{Spin-Supercurrent via Magnetic Insulators}

\begin{figure}[b!]
\centering
   \includegraphics[width=1\linewidth]{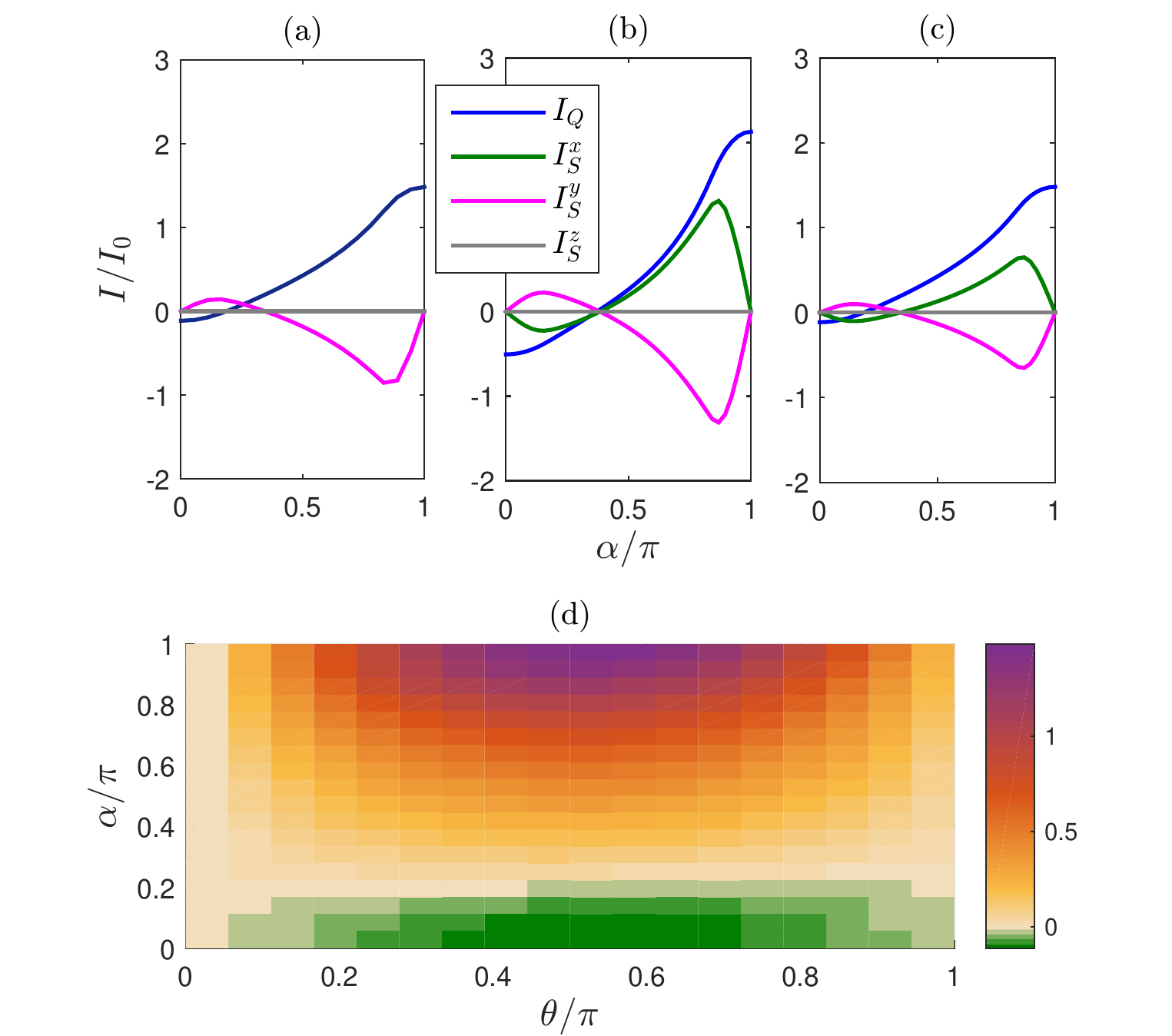}
   \caption{(Color online) Plot of the spin- and charge-supercurrents in the system. We have used $\xi_S=30$ nm and the relative temperature $T/T_c=0.02$. The interface parameters are set to be equal, $G_\varphi=3$ and $\zeta=2$, and the phase difference is the one supporting the critical current, $\theta=\pi/2$. In (a), we have set $d=20$ nm, $\phi=0$. In (b), we have $d=5$ nm and $\phi=\pi/4$. In (c), we set $d=20$ nm and $\phi=\pi/4$. As expected, the components of the spin-supercurrents are mirror-images of each other in (b) and (c) due to the choice of magnetic configuration of the insulators, $\phi=\pi/4$ [see Eq. (\ref{eq:phi})]. The charge-supercurrent is independent of $\phi$. As seen, it changes sign when going from $\alpha=0$ to $\alpha=\pi$, signalling a 0-$\pi$ transition. The normalization constant of the charge-current is $I_0 = N_0eDA/4$ while for the spin-currents it is $I_0 = N_0\hbar DA/8$. The contour plot in the bottom panel (d) is the charge-supercurrent in the $\theta$-$\alpha$ plane using $d=20$ nm, showing the occurrence of the 0-$\pi$ transition around $\alpha\simeq 0.2\pi$ (the dark green region corresponds to the $\pi$-phase).}
\label{fig:current} 
\end{figure}

We proceed to discuss how the charge- and spin-supercurrents sustained by the system are influenced by the presence of the ferromagnetic insulators. In the quasiclassical framework, these are given by 
\begin{align}
I_Q = \frac{N_0eDA}{4} \int^\infty_{-\infty} \text{d}\varepsilon \text{Tr}\{ \rho_3 (\check{g}\partial_x\check{g})^\text{K} \}
\end{align}
and 
\begin{align}
I_S^\nu = \frac{N_0\hbar DA}{8} \int^\infty_{-\infty} \text{d}\varepsilon \text{Tr}\{ \rho_3\tau_\nu (\check{g}\partial_x\check{g})^\text{K} \}.
\end{align}
Here, $N_0$ is the density of states at the Fermi-level in the normal-state, $e$ is the electric charge, $\hbar$ is the reduced Planck constant, while $A$ is the interface contact area. For future use, we also define the bulk superconducting coherence length $\xi_S=\sqrt{D/\Delta_0}$. In the weak proximity effect regime, we were able to find a general analytical result for the supercurrents of spin and charge (see Appendix for details). We first briefly consider the charge-supercurrent which reads:
\begin{align}
I_Q = (I_{Q,0} + I_{Q,1}\cos\alpha G_{\varphi}^L G_\varphi^R)\sin\theta,
\end{align}
where the coefficients $I_{Q,0}$ and $I_{Q,1}$ are lengthy expressions that depend on junction parameters such as the width $d$, misalignment angle $\alpha$, temperature $T$, and the interface transparencies $\zeta_{L/R}$. The charge-current is found to be independent of which orientation the magnetic moments have in the $xy$-plane, $\phi$. We see that the presence of magnetic insulators coupled to the superconductors introduces a $\cos\alpha$-dependence on the supercurrent, not only tuning its overall magnitude, but also changing the quantum ground-state of the junction from 0 to $\pi$ when 
\begin{align}
I_{Q,1}\cos\alpha G_{\varphi}^L G_\varphi^R = -I_{Q,0}.
\end{align}
Thus, 0-$\pi$ transitions can now occur even with a normal metal interlayer by changing $\alpha$, a feature which was also reported in the ballistic limit in Ref. \cite{zaikin_prb_09}. To demonstrate that this is a robust feature, we have computed the charge supercurrent without any assumption of a weak proximity effect, thus using the full Riccati parametrization. This is shown in Fig. \ref{fig:current}(d), where the current changes sign at $\alpha\simeq 0.2\pi$ corresponding to the 0-$\pi$ transition. Further information may be inferred from the analytical expression for the charge-supercurrent in the Appendix, Eq. (\ref{eq:charge}): as the width $d$ of the junction increases, larger values for the spin-dependent phase-shifts $G_\varphi$ are required in order to make the 0-$\pi$ transition possible.

\begin{figure}[b!]
\centering
\resizebox{0.52\textwidth}{!}{
\includegraphics{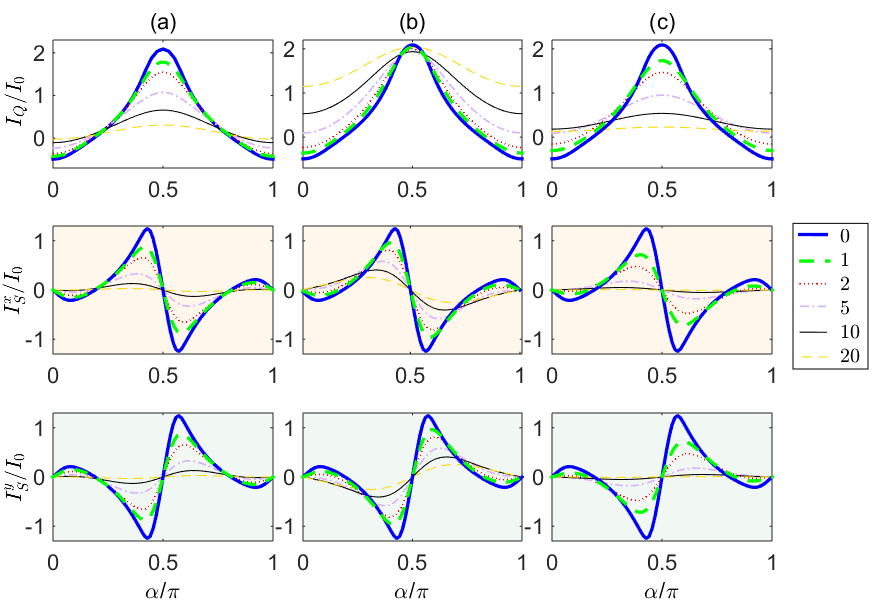}}
\caption{(Color online) Plot of the charge-supercurrent and the components of the spin-supercurrent in the presence of spin-flip scattering due to magnetic impurities and spin-orbit impurity scattering. Column (a) corresponds to magnetic impurity scattering (lines corresponding to different values of $g_\text{sf}$), column (b) to spin-orbit scattering ($g_\text{so}$), and column (c) to both present with equal magnitude. The parameters are set to $d = 5$ nm, $\xi_S = 30$ nm, $G_\varphi = 3$, $\zeta = 2$, $T/T_c = 0.02$, and $\theta=\pi/2$}
\label{fig:Current_flip} 
\end{figure}

Interestingly, there exists not only a superflow of charge in the system, \textit{but also of spin}. The polarization occurs in the direction $\mathbf{m}_L\times\mathbf{m}_R$, and so we find that while $I_S^z=0$, one has:
\begin{align}\label{eq:main}
I_S^x &= G_\varphi^L G_\varphi^R \sin\phi\sin\alpha(I_{S,0} + I_{S,1}\cos\theta).
\end{align}
Eq. (\ref{eq:main}) is one of the main results of this work. It is seen that the spin-supercurrent vanishes if one only has one magnetic insulator, in which case $G_\varphi^L$ or $G_\varphi^R$ is zero. Moreover, it is proportional to $\sin\alpha$, which shows that it is also absent in the P or AP alignment $(\alpha=0,\pi)$. For other angles $\alpha$, however, it is in general present. The coefficients $\{I_{S,0},I_{S,1}\}$ are purely real and vanish in the absence of superconductivity ($\Delta=0$). There exists a simple relation between the components of the spin-supercurrent in the $xy$-plane:
\begin{align}\label{eq:phi}
\frac{I_S^x}{I_S^y} = -\frac{\sin\phi}{\cos\phi}.
\end{align}
This spin-supercurrent has several remarkable features: first of all, it is conserved throughout the normal metal just like the charge-current. Secondly, it is long-ranged as it flows through a normal metal without any exchange field. Thirdly, it has one component that is \textit{independent} of the superconducting phase difference $\theta$. The other component goes like $\cos\theta$, meaning that the total spin-supercurrent satisfies $I_S^x(\theta) = I_S^x(-\theta)$. This can be understood physically, since a spin-current is invariant under time-reversal symmetry. The latter operation transforms $\theta \to (-\theta)$ and causes the charge-supercurrent to change sign. In the Appendix, we give the full expression of the spin-supercurrents including the coefficients and their dependence on the junction parameters. The spin-supercurrent vs. misalignment angle $\alpha$ is shown in Fig. \ref{fig:current}(a)-(c) for different junction parameters.

The spin-supercurrent should be possible to control with a weak external field coupling to the magnetic insulators, tuning their relative orientation. We underline that our structure, unlike previous works, does not include any ferromagnetic metals. In fact, a conceptually similar experimental structure to the one that we propose to use in our manuscript has recently been demonstrated in Ref. \cite{li_prl_13}. There, the authors investigated a spin-valve structure consisting of a superconductor flanked by two magnetic insulators of slightly different thicknesses (both of order a few nm). Applying an external magnetic field would then control the magnetization orientation of the thinner of these layers. This indicates that our results are of experimental relevance using currently available techniques. 

In an experimental setting, the normal metal sample may well include some degree of magnetic impurities or spin-orbit scattering on impurities. Thus, it is of interest to see how such non-ideal effects influence the predictions made in this paper. Although no tractable analytical expression is accessible in this case, we have computed numerically the charge- and spin-supercurrent in the presence of spin-flip scattering and spin-orbit impurity scattering as described by Eq. (\ref{eq:flip}). Interestingly, the dissipationless flow of charge and spin are affected very differently depending on the type of scattering. Consider first the charge-supercurrent [top row of Fig. \ref{fig:Current_flip}]. With increasing spin-flip scattering, the current is monotonically suppressed. However, this is not the case for spin-orbit impurity scattering (middle panel). Instead, the 0-$\pi$ transition point vanishes and the current retains its order of magnitude even for very large values of $g_\text{so}$. Turning to the spin-supercurrent, we find that both magnetic impurity scattering and spin-orbit impurity scattering suppress the spin-flow monotonically.

The physical origin of the different behavior of the charge- and spin-supercurrents when adding magnetic impurities and spin-orbit scattering can be traced back to how the singlet and triplet superconducting correlations are affected by them \cite{faure_prb_06, linder_prb_08}. It can be demonstrated analytically that the singlet component is insensitive to spin-orbit impurity scattering in the absence of a magnetic field, as is reasonable since spin-orbit scattering respects time-reversal symmetry. On the other hand, the triplet component is suppressed as the spin-orbit scattering rate increases. Based on this, one can now understand why the charge-supercurrent evolves different with increasing spin-flip and spin-orbit scattering respectively. In the former case, both the singlet and triplet components are suppressed, i.e. the total superconducting proximity effect is reduced, and the current simply decays montonically. In the latter case, however, only the triplet part is suppressed. With only the singlet part remaining in the normal metal, there is no mechanism to cause a 0-$\pi$ transition and the sign of the current remains positive and still of appreciable magnitude. 

The same reasoning can be applied to the spin-current. In this case, it is solely the triplet part which is responsible for its existence. Since both magnetic impurities and spin-orbit scattering suppress the triplet Cooper pairs, the spin-supercurrent decays monotonically as the scattering rate increases. It can be seen from Fig. \ref{fig:Current_flip} that the dependence of the spin-supercurrent on the misalignment angle $\alpha$ between the magnetic insulator moments goes toward a pure $\sin\alpha$ profile as $g_\text{so}$ becomes large.

\begin{figure}[b!]
\centering
   %\begin{subfigure}[b]{0.50\textwidth}
   \includegraphics[width=1\linewidth]{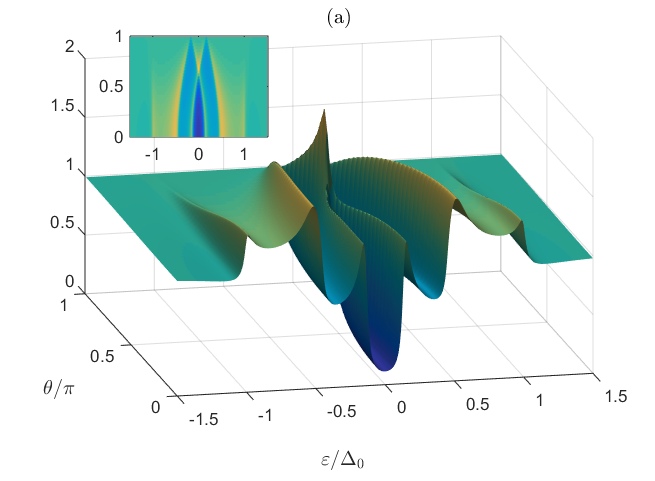}
  % \caption{}
%\end{subfigure}
%\begin{subfigure}[b]{0.50\textwidth}
   \includegraphics[width=1\linewidth]{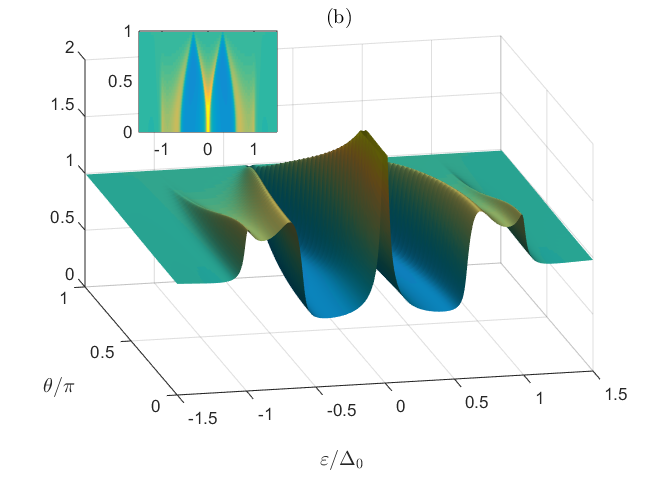}
  % \caption{}
%\end{subfigure}
%\begin{subfigure}[b]{0.50\textwidth}
   \includegraphics[width=1\linewidth]{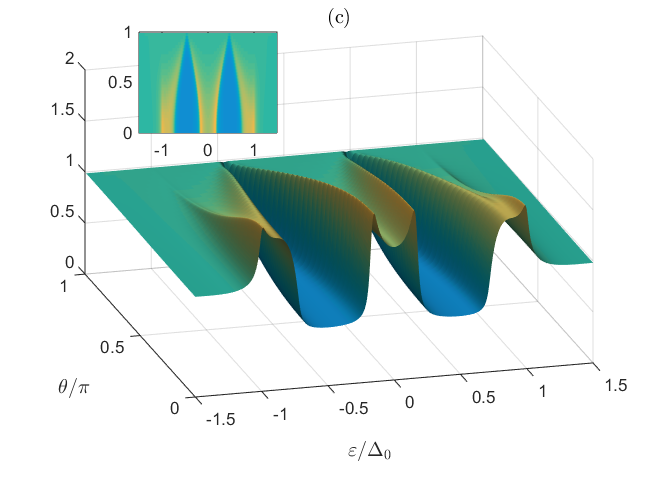}
   %\caption{}
%\end{subfigure}
\centering
\caption{(Color online) Proximity-induced density of states $N(\varepsilon,\theta)$ (normalized to its normal-state value) in the middle of the normal metal region. In all cases, we are considering the P configuration where both magnetic insulators have moments pointing in the $z$-direction. The strength of the spin-dependent phase-shifts occuring at the interfaces are given by (a) $G_\varphi=0.55$, (b) $G_\varphi=1.05$, and (c) $G_\varphi=1.55$. We have used the parameters $d=20$ nm, $\xi_S=30$ nm, and $\zeta=5$.}
   \label{fig:dos} 
\end{figure}

\subsection{Phase-tunable Triplet Superconductivity}
Besides the appearance of this unusual spin-supercurrent, a Josephson junction with magnetic insulator interfaces offers a unique way to control triplet superconductivity as we now demonstrate. It has previously been shown that a crossover from pure conventional even-frequency pairing to odd-frequency pairing is made possible via spin-active interfaces in S$\mid$MI$\mid$N bilayer junctions \cite{linder_prb_10}. The pivotal parameter is the ratio between the spin-dependent phase-shifts and the normal-state tunnel conductance, in our notation $G_\varphi$, which causes a pure odd-frequency proximity pairing state at the Fermi level ($\varepsilon=0$) when $G_\varphi>1$ while a pure even-frequency state occurs when $G_\varphi<1$. Experimentally, this is manifested as a large zero-energy peak in the density of states when the odd-frequency triplets dominate. Conversely, a minigap appears in the spectrum for conventional singlet pairing. To observe such an effect, it would be necessary to fabricate several samples with a varying ratio $G_\varphi$. One way to accomplish this could be to vary the width of the MI interlayer in order to tune the tunneling probability. 

Instead, we here show that when spin-active interfaces are incorporated in a Josephson junction geometry, the crossover from even- to odd-frequency pairing can now be controlled by the \textit{superconducting phase difference} $\theta$, which in turn is determined by the current flowing through the system. This offers a new way to induce a triplet proximity effect which can be changed \textit{in situ} within a single sample, simply by varying $\theta$. The crossover is manifested by the qualitative nature of the proximity effect, going from a minigap (conventional singlet proximity effect) to an enhanced low-energy peak (odd-frequency triplet proximity effect). In order to probe how the change in pairing symmetry is manifested experimentally, we here compute the density of states in the normal metal region and its phase-dependence numerically which allows us to relax the assumption of a weak proximity effect. The DOS normalized to its normal-state value is obtained from the solution of the Ricatti equations via:
\begin{align}
N(\varepsilon,\theta) = \text{Re}\{\text{Tr}[\N(\underline{1}+\g\gt)]\}/2.
\end{align}
To make better contact with experimental measurements, we have added a small imaginary part to the quasiparticle energies, $\varepsilon \to \varepsilon + \i\delta$ where $\delta/\Delta_0\ll1$, which represents inelastic scattering.

We consider the most general case where each superconducting interface contains a magnetic insulator. The results are shown in Fig. \ref{fig:dos}, where we have focused on the experimentally most accessible configuration with the insulators in the P state. Three choices of the strength of the spin-active scattering at the interfaces are considered in (a)-(c) with $G_\varphi=\{0.55, 1.05, 1.55\}$. It is seen that the nature of the superconducting proximity changes qualitatively due to the presence of the magnetic insulators. It is known that in the absence of magnetic elements $(G_\varphi=0)$, a minigap is induced in the normal metal which is largest for $\theta=0$ and closes at $\theta=\pi$. In Fig. \ref{fig:dos}(a), the minigap is prominent at small phase-differences $\theta$. However, instead of monotonically closing the minigap as $\theta$ is driven towards $\pi$, the density of states becomes strongly enhanced at low energies. This feature arises due to the odd-frequency symmetry of the triplet Cooper pairs in the normal metal \cite{tanaka_prl_07, linder_prb_10, kawabata}. When the spin-active scattering, taking place at the insulators, becomes stronger in Fig. \ref{fig:dos}(b) and (c), the minigap has vanished all-together, leaving behind only a clear zero-energy peak in the density of states. 

\begin{figure}[b!]
\centering
\resizebox{0.5\textwidth}{!}{
\includegraphics{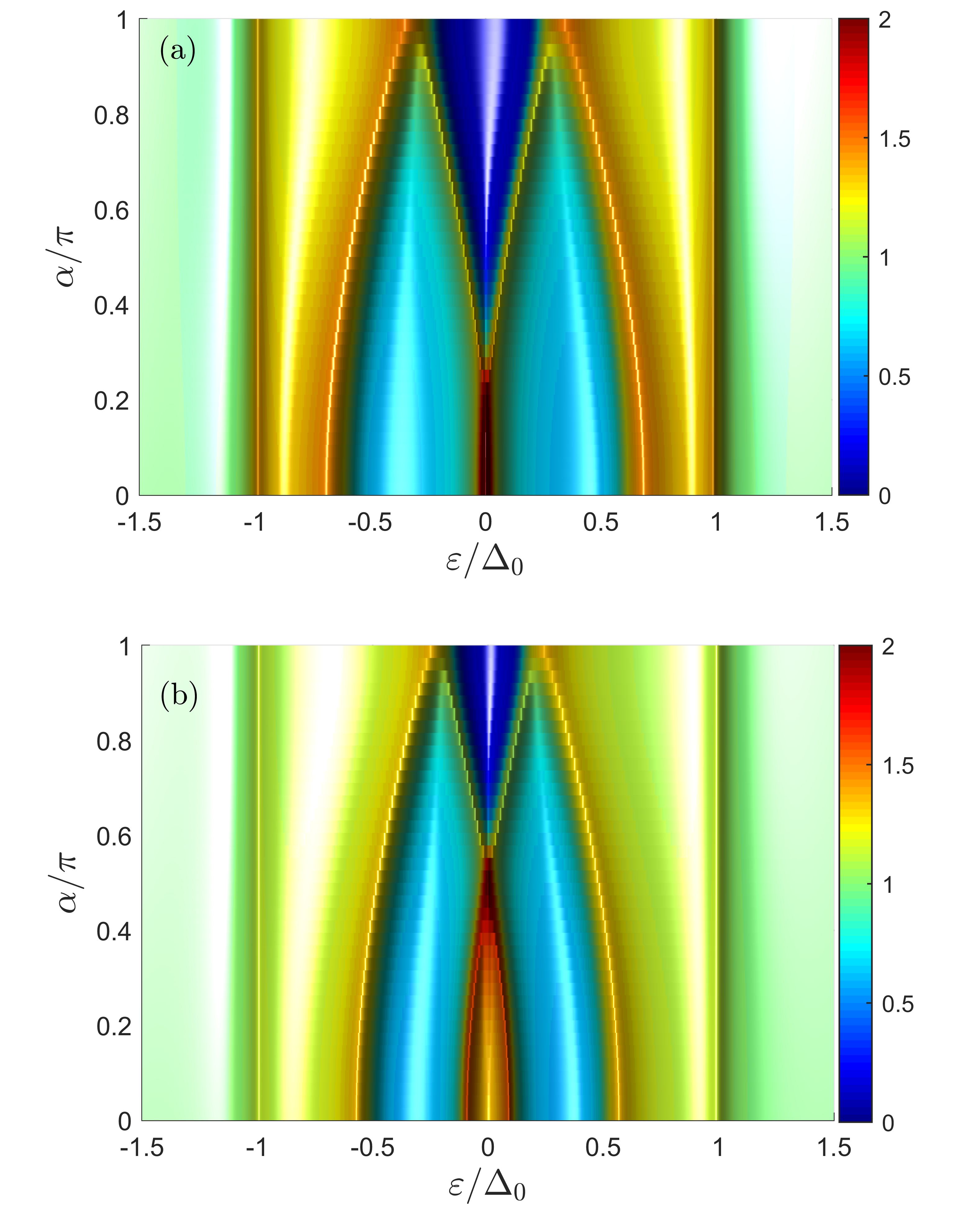}}
\caption{(Color online) Proximity-induced density of states $N(\varepsilon,\theta)$ (normalized to its normal-state value) in the middle of the normal metal as a function of quasiparticle energy $\varepsilon$ and the misalignment angle $\alpha$ of the magnetic insulators. In (a), the superconducting phase bias is set to $\theta=0$, corresponding to zero current-flow. In (b), we have $\theta=\pi/2$, corresponding to the critical current-flow. The other parameters are set to $d=20$ nm, $\xi_S=30$ nm, $G_\varphi=1.05$, $\zeta=5$. }
\label{fig:dos2} 
\end{figure}

The qualitative change in the density of states (going from fully suppressed to enhanced at low energies) can be seen clearly also when keeping the superconducting phase difference $\theta$ fixed and varying the magnetic configuration $\alpha$. This is shown in Fig. \ref{fig:dos2}: as one changes from a P to AP configuration (going from $\alpha=0$ to $\alpha=\pi$), the system makes a transition from hosting proximity-induced triplet superconductivity to singlet superconductivity. Our work thus demonstrates a \textit{conversion between singlet and triplet Cooper pairs in a normal metal by tuning either the superconducting phase difference or the configuration of the magnetic insulators}. This has the important advantage that it can be done \textit{in situ}, as opposed to using ferromagnets where \eg several samples with different widths are created to suppress the singlet component relative the triplet one.

\begin{figure}[t!]
\centering
\resizebox{0.48\textwidth}{!}{
\includegraphics{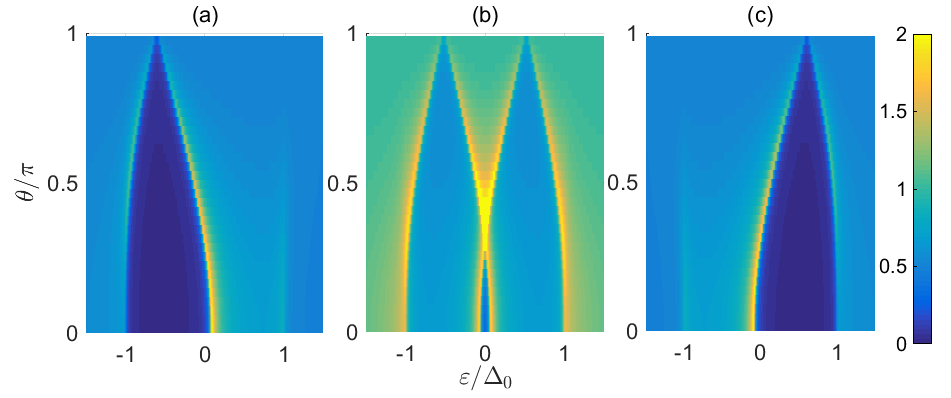}}
\caption{(Color online) Spin-resolved density of states (normalized to its normal-state value) in the $\varepsilon-\theta$ plane. In (b), the total density of states is shown whereas in (a) and (c) the spin-$\downarrow$ and spin-$\uparrow$ contributions are shown, respectively. The parameters used are $\zeta=5$, $G_\varphi=0.9$, $d=10$ nm, $\xi_S = 30$ nm, and $\alpha=0$.}
\label{fig:spin} 
\end{figure}

It is interesting to note that the density of states for each electron-spin is highly non-degenerate and tunable, as shown in Fig. \ref{fig:spin}. This could potentially be utilized in creating large thermoelectric effects based on the idea of Ref. \cite{machon_prl_13} which demonstrated that the spin-splitted density of states arising in superconductor/ferromagnet hybrids could yield a thermoelectric figure of merits far exceeding what is obtained in the non-superconducting phase.

\begin{figure}[b!]
\centering
\resizebox{0.5\textwidth}{!}{
\includegraphics{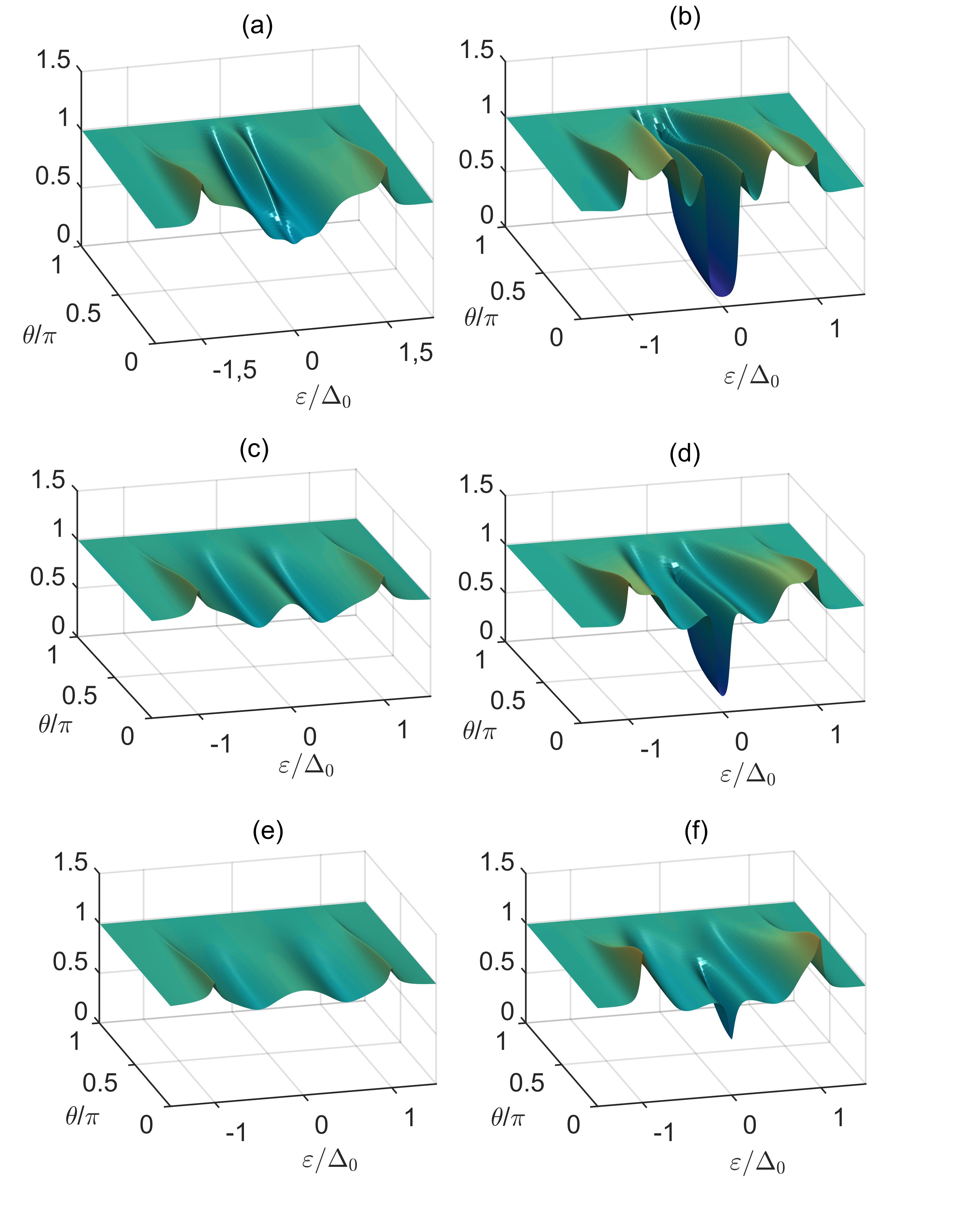}}
\caption{(Color online) Plot of the density of states $N(\varepsilon,\theta)$ (normalized to its normal-state value) in the energy-superconducting phase difference ($\varepsilon$-$\theta$) plane. The top row illustrates the case where $G_\varphi=0.55$ and (a) $g_\text{sf}=0.05$, (b) $g_\text{so}=0.05$. The middle row has $G_\varphi=1.05$ and (c) $g_\text{sf}=0.10$, (d) $g_\text{so}=0.10$. The bottom row shows $G_\varphi=1.55$ and (e) $g_\text{sf}=0.15$, (f) $g_\text{so}=0.15$. The remaining parameters are set to $d=20$ nm, $\xi_S=30$ nm, and $\zeta=5$, in the P configuration.}
\label{fig:DOS_flip} 
\end{figure}

Similarly to our treatment of the spin- and charge-supercurrent, we also investigate the influence of spin-flip and spin-orbit scattering on the density of states. In the left column of Fig. \ref{fig:DOS_flip}, we consider different values for $G_\varphi$ and the spin-flip scattering rate. Regardless of the value of $G_\varphi$, in particular of whether it is smaller than or greater than the critical value $G_{\varphi,c}=1$, the influence of the superconducting proximity effect on the DOS is diminished. As discussed previously in the context of the charge- and spin-supercurrents, this may be understood physically from the fact that magnetic impurities suppress singlet and triplet components alike, such that the DOS eventually reverts to its normal-state value for any phase-difference and energy. The situation is different when considering spin-orbit impurity scattering, shown in the right column of Fig. \ref{fig:DOS_flip}. Now, the spectroscopic manifestation of the superconducting proximity effect depends on whether we are in the singlet-dominated regime $G_\varphi<1$ or the triplet-dominated regime $G_\varphi>1$. For $G_\varphi=0.55$, the presence of spin-orbit scattering leaves the minigap intact while suppressing the zero-energy peak that emerges as the superconducting phase-difference increases. Hence, the superconducting proximity effect remains clearly visible in the DOS. For $G_\varphi>1$ shown in (d) and (f), however, increasing the spin-orbit scattering rate causes the low-energy enhancement of the DOS to be absent since the triplet component is suppressed by this type of scattering. When applying even stronger values of spin-orbit scattering, a clear minigap appears for all the values of $G_\varphi$. It should also be noted that the spectroscopic fingerprints of the superconducting proximity effect are much more sensitive toward the presence of magnetic impurity and spin-orbit scattering than the charge- and spin-supercurrents, the former being suppressed in magnitude faster compared to the current at a given value of $g_\text{sf/so}$.

\section{Discussion and concluding remarks}

We here discuss in more detail how our work is related to previous findings. In the proposal by Houzet and Buzdin \cite{houzet_prb_07}, a ferromagnetic trilayer was suggested as the minimal structure that would be able to generate a long-ranged triplet supercurrent.  In another work by Grein \etal\text{ }\cite{grein_prl_09}, strongly polarized ferromagnets with two spin-active interfaces were considered, thus in some sense being similar to the trilayer system of Ref. \cite{houzet_prb_07} with the exception that the spin-bands were now assumed to be completely decoupled in the bulk due to the large exchange field. In this case, a spin-supercurrent was shown to also be generated.  Shomali \etal\text{ }\cite{shomali_njp_11} studied the spin-current in a Josephson junction with a ferromagnetic metal bilayer and it was realized that a long-ranged supercurrent in ferromagnets could in fact be generated with only two ferromagnets \cite{trifunovic_prl_10}, albeit only as a higher-order effect. More precisely, there would be a contribution to a long-ranged triplet supercurrent from the second Josephson harmonic $\sin(2\theta)$. This could make experimental detection difficult, since the magnitude of the second harmonic latter is usually much smaller than the first harmonic, and a very specific fine-tuning of the junction parameters would be required to observe the effect. Spin-supercurrents have also been analyzed in other types of superconducting structures, including magnetic textures such as spirals, and also using intrinsically triplet bulk superconductors \cite{gronsleth, brydon, kulic, nogueira, annunziata}. Very recently \cite{moor_sst_15}, spin supercurrents in junctions composed of multiband superconductors coexisting with a spin-density wave state was studied theoretically. Similar dependencies on the superconducting phase difference and magnetic misalignment between the spin-density waves as in our case was shown, even if the system under consideration in this work is physically quite different from Ref. \cite{moor_sst_15}. 

The spin-supercurrent reported in our work occurs in the first harmonic, i.e. it is not a higher-order effect, meaning that it is present without any fine-tuning of parameters in order to suppress the first harmonics in favor of higher ones. Moreover, it occurs \textit{without use of any ferromagnetic metals}: the spin superflow takes place in a non-magnetic normal metal. There are several different choices for magnetic insulators that can be used in the proposed setup shown in Fig. \ref{fig:model}. Previous experiments considering superconducting hybrid structures have utilized magnetic insulators such as EuO \cite{tedrow_prl_86}, EuS \cite{li_prl_13}, and GdN \cite{pal_ncom_14}. The particular choice of magnetic insulator also depends on how well it can be grown at the interface between the superconductor and the normal metal. We speculate that suitable material combinations to construct our setup could be Nb and EuO as the superconductor and magnetic insulator, or alternatively NbN and GdN. Concerning the phase-dependent density of states in the normal metal, experimental techniques are available for measuring this quantity as demonstrated in Ref. \cite{sueur_prl_08} in a conventional SNS Josephson junction. By integrating the junction in a loop geometry, the superconducting phase $\theta$ is then tunable via a minute magnetic flux. Using AFM-spectroscopy, a complete mapping of how the density of states evolves spatially through the junction as a function of $\theta$ is possible. 

We note that very recently, quasiclassical boundary conditions valid for \textit{any} strength of the barrier polarization were derived \cite{eschrig_njp_15}. This opens the possibility to study theoretically systems with very strongly spin-polarized magnetic insulators and even half-metallic (fully polarized) ferromagnets. Concerning direct experimental detection of the spin-polarization of the supercurrent itself, one possibility could be to, once it has been created, inject it into a magnetic material with a weak magnetic anisotropy and observe the resulting magnetization dynamics as a result of spin-transfer. There have also been proposals utilizing optical detection of spin transport through non-magnetic metals \cite{fohr} as well as electrical detection \cite{valenzuela} via a spin-current induced Hall effect. Further work is needed to clarify precisely if and how this could be possible using spin-supercurrents in superconducting structures.

In summary, we have shown that by integrating superconductors with magnetic insulators, one arrives at a unique way to both create and control triplet superconductivity in a well-defined way with the superconducting phase-difference, and to also create a conserved and tunable spin-supercurrent flowing through a normal metal. 

\begin{center}
\textbf{Acknowledgments}
\end{center}
The authors thank S. Jacobsen, J. A. Ouassou, I. Kulagina, A. Pal, A. di Bernardo, J. Robinson, M. Blamire, and M. Eschrig for many useful discussions. This work was supported by the "Outstanding Academic Fellows" programme at NTNU, the
COST Action MP-1201 'Novel Functionalities through Optimized
Confinement of Condensate and Fields’, and Norwegian Research Council grants no. 205591 and no. 216700.
\begin{widetext}
\section{Appendix: Detailed expressions for charge- and spin-supercurrents}

We here provide comprehensive results for the analytical expressions of the supercurrents of charge and spin supported by the system. In the weak proximity effect, one finds the following completely general expressions:
\begin{align}
I_Q &= \frac{N_0eDA}{4}\int^{\infty}_0 \text{d}\varepsilon \text{tanh}\Big(\frac{\beta\varepsilon}{2}\Big) 4 \text{Re}\Big\{[ 2f_s\partial_x \tilde{f}_s - 2f_t\partial_x\tilde{f}_t - f_\uparrow\partial_x \tilde{f}_\uparrow - f_\downarrow\partial_x\tilde{f}_\downarrow]- [\tilde{\ldots}]  \Big\},\notag\\
I_S^x &= \frac{N_0\hbar DA}{8}\int^{\infty}_0 \text{d}\varepsilon \text{tanh}\Big(\frac{\beta\varepsilon}{2}\Big) 4 \text{Re}\Big\{[-(f_\uparrow+f_\downarrow)\partial_x \tilde{f}_t - f_t\partial_x(\tilde{f}_\uparrow+\tilde{f}_\downarrow)] - [\tilde{\ldots}] \Big\},\notag\\
I_S^y &= \frac{N_0\hbar DA}{8} \int^{\infty}_0 \text{d}\varepsilon \text{tanh}\Big(\frac{\beta\varepsilon}{2}\Big) 4 \text{Im}\Big\{ [ (f_\uparrow-f_\downarrow)\partial_x \tilde{f}_t - f_t\partial_x (\tilde{f}_\uparrow-\tilde{f}_\downarrow)] + [\tilde{\ldots}] \Big\},\notag\\
I_S^z &=  \frac{N_0\hbar DA}{8} \int^{\infty}_0 \text{d}\varepsilon \text{tanh}\Big(\frac{\beta\varepsilon}{2}\Big) 4 \text{Re}\Big\{ [-f_\uparrow\partial_x \tilde{f}_\uparrow + f_\downarrow\partial_x\tilde{f}_\downarrow] - [\tilde{\ldots}] \Big\}.
\end{align}
Above, the notation $(\tilde{\ldots})$ means changing the sign of energy and complex conjugate and we defined the inverse temperature $\beta=1/(k_BT)$. It is seen that the spinless singlet correlations $f_s$ do not contribute to any of the spin-currents. In the special case of a normal metal separating the superconductors, one can work further with the above expressions by inserting the solutions
\begin{align}
f_m = A_m\e{\i kx} + B_m\e{-\i kx}, m=\{s,t,\uparrow,\downarrow\}.
\end{align} We then get expressions for the supercurrents which is independent of position:
\begin{align}
I_Q &= N_0eDA\int^{\infty}_0 \text{d}\varepsilon \text{tanh}\Big(\frac{\beta\varepsilon}{2}\Big)  \text{Re}\Big\{ 2\i k[(\Au\tAu - \Bu\tBu) + 2(A_t\tAt - B_t\tBt) - 2(A_s\tAs - B_s\tBs)+ (\Ad\tAd - \Bd\tBd)]  \Big\}, \notag\\
I_S^x &= \frac{N_0\hbar DA}{2}\int^{\infty}_0 \text{d}\varepsilon \text{tanh}\Big(\frac{\beta\varepsilon}{2}\Big)   \text{Re}\Big\{ 2\i k[(\Au+\Ad)\tAt - (\Bu+\Bd)\tBt +(\tAu+\tAd)A_t - (\tBu+\tBd)B_t] \Big\}, \notag\\
I_S^y &= \frac{N_0\hbar DA}{2}\int^{\infty}_0 \text{d}\varepsilon \text{tanh}\Big(\frac{\beta\varepsilon}{2}\Big)   \text{Re}\Big\{2k[-(\Au-\Ad)\tAt+(\Bu-\Bd)\tBt +(\tAu-\tAd)A_t-(\tBu-\tBd)B_t] \Big\}, \notag\\
I_S^z &= \frac{N_0\hbar DA}{2}\int^{\infty}_0 \text{d}\varepsilon \text{tanh}\Big(\frac{\beta\varepsilon}{2}\Big)  \text{Re}\Big\{ 2\i k[(\Au\tAu - \Bu\tBu) - (\Ad\tAd - \Bd\tBd)] \Big\}.
\end{align}
The coefficients $\{A_m,B_m\}$ for the singlet and each of the triplet components are determined by the boundary conditions. For instance, one finds for the charge-supercurrent that
\begin{align}\label{eq:charge}
I_Q &=N_0eDA\sin\theta\int^{\infty}_0 \text{d}\varepsilon \text{tanh}\Big(\frac{\beta\varepsilon}{2}\Big)\text{Re}\Big\{ 4\i k \Gamma^{-1} \sin(kd)\sinh^2\Theta \Big(k^2d^2\zeta_L\zeta_R + G_{\varphi}^L G_{\varphi}^R \cos\alpha\Big)\Big\}.
\end{align}
upon defining the quantity:
\begin{align}
\Gamma &= \Big(k^2 d^2 \zeta_R^2 + (G_{\varphi}^R)^2\Big)\Big(k^2d^2\zeta_L^2-(G_{\varphi}^L)^2 + 2(G_{\varphi}^L)^2\cos^2\alpha \Big)\cos^2(kd) - \Big(k^2d^2\zeta_L\zeta_R - G_{\varphi}^L G_{\varphi}^R\cos\alpha\Big)^2,\notag\\
\end{align}
 % alt $(L^2k^2\zeta_R^2+(G_{\phi}^R)^2)(L^2k^2\zeta_L^2+2(G_{\phi}^L)^2\cos^2\alpha-(G_{\phi}^L)^2)\cos^2(kL)-(L^2k^2\zeta_L\zeta_R-G_{\phi}^L G_{\phi}^R\cos\alpha)^2$
The spin-supercurrent is given by: 
\begin{align}
I_S^x &= \frac{N_0\hbar DA\sin\phi\sin\alpha G_\varphi^L G_\varphi^R }{2}\int^{\infty}_0 \text{d}\varepsilon \text{tanh}\Big(\frac{\beta\varepsilon}{2}\Big)\text{Re}\Big\{ 4\i k \Gamma^{-2} \sin(kd)\sinh^2\Theta \Big(a_1 + a_2 \cos\theta\Big)\Big\},
\end{align} where we have defined the expressions

\begin{align}
a_1 &= \Big(2(G_{\varphi}^L)^2\cos^2\alpha - (G_{\varphi}^L)^2+k^2d^2(\zeta_L^2+\zeta_R^2)+(G_{\varphi}^R)^2\Big)\Big(k^2d^2\zeta_L\zeta_R - G_{\varphi}^L G_{\varphi}^R\cos\alpha\Big)\cos^2(kd),\\
a_2 &= \Big(k^2d^2\zeta_L^2+2(G_{\varphi}^L)^2\cos^2\alpha - (G_{\varphi}^L)^2\Big)\Big(k^2d^2\zeta_R^2+(G_{\varphi}^R)^2\Big)\cos^2(kd)+\Big(k^2d^2\zeta_L\zeta_R - G_{\varphi}^L G_{\varphi}^R\cos\alpha\Big)^2.
\end{align}

\end{widetext}

\end{document}